\documentclass[12pt]{article}
\begin{document}
\title{FUZZY SPACETIME:FURTHER CONSIDERATIONS}
\author{B.G. Sidharth\\
International Institute for Applicable Mathematics \& Information Sciences\\
Hyderabad (India) \& Udine (Italy)\\
B.M. Birla Science Centre, Adarsh Nagar, Hyderabad - 500 063 (India)}
\date{}
\maketitle
\begin{abstract}
We consider fuzzy spacetime, quanta of area and related concepts in the context of latest approaches to Quantum Gravity and show its interface with usual non-Abelian gauge theory. We also discuss in this context a cosmology which correctly predicted a dark energy driven acceleratling universe with a small cosmological constant, amongst other things.
\end{abstract}
\section{Introduction}
Inspite of several fruitless decades of work, the two great intellectual pillars of twentieth century Physics, General Relativity and Quantum Theory have remained irreconcible though each has proved to be successful in its own domain. As Wheeler put it \cite{mwt} the problem is the introduction of spin half into General Relativity and curvature into Quantum Theory. On the other hand, inspite of the successes of either theory, there are still a number of unresolved problems and unless these are addressed, a complete or unified description may not be possible. Thus there is the question of spacetime singularities termed by Wheeler to be the Greatest Crisis of Physics or as yet directly undetected gravitational waves on the one hand, and on the other, from the domain of Particle Physics, the existence of some eighteen arbitrary parameters in the Standard Model, or the elusive monopoles, the puzzle of the muon $g$ factor or the tiny mass of the neutrino as brought out by the Superkamiokande experiment and the characterization of the newly determined dark energy and so on. Even leaving these considerations aside, the unsatisfactory character of the short distance behavior of the Standard Model itself points to Physics beyond the Standard Model. In the words of 't Hooft \cite{gerard}, ``...the standard model as it stands today cannot be entirely correct... a reason must be found as to why the forces at short time scales balance out. The way things are for the elementary particles, at present, is that the forces balance out just by accident. It would be an inexplicable accident, and as no other examples of such accidents are known in Nature, at least not of this magnitude, it is reasonable to suspect that the true short distance structure is not exactly as described in the Standard Model...''.\\
It is noteworthy that the two desparate realms of General Relativity and Quantum Theory nevertheless share a common feature: They operate within a differenciable spacetime manifold viz., Reimannian spacetime and Lorentzian spacetime respectively. Over the past few decades some progress towards the Quantum Theory of Gravity has been made by discarding the concept of smooth spacetime, be it in Quantum SuperString Theory or other Quantum Gravity approaches, as also the one to be described below. In these schemes there is a minimum spacetime cut off at the Planck or more generally the Compton scale. Indeed to quote 't Hooft himself \cite{hooft}, ``It is somewhat puzzling to the present author why the lattice structure of space and time had escaped attention from other investigators up till now...''.
\section{The Minimum Cut Off}
To fix physical ideas let us start with the Kerr-Newman metric, applied to the electron \cite{cu}. This purely classical description, amazingly enough, heals the purely Quantum Mechanical $g = 2$ factor of the electron, though there is now a naked singularlity, in that the horizon becomes complex:
\begin{equation}
r_+ = \frac{GM}{c^2} + \imath b, b \equiv (\frac{G^2Q^2}{c^8} + a^2 - \frac{G^2M^2}{c^4})^{1/2}\label{e1}
\end{equation}
$G$ being the gravitational constant, $M$ the mass and $a \equiv L/Mc,L$ being the angular momentum.\\
On the other hand the position coordinate of the electron from the Dirac theory \cite{dirac} is also given by a complex coordinate:
\begin{equation}
x = (c^2p_1H^{-1}t) + \frac{\imath}{2} c\hbar (\alpha_1 - cp_1H^{-1})H^{-1}\label{e2}
\end{equation}
Moreover the imaginary parts in either case (\ref{e1}) and (\ref{e2}) have the same order, that of the electron Compton wavelength. As Dirac argued, Quantum Mechanical measurements of spacetime points imply infinite energies and momentum. Meaningful Physics is recovered only on averaging over intervals of the order of the Compton scale, in which case the complex part of the coordinate disappears (Cf.ref.\cite{dirac}). Indeed wlithin the Compton scale, unphysical zitterbewegung and superluminal effects are encountered.\\
When minimum spacetime intervals are introduced, we immediately have a non-commutative geometry, as shown by Snyder \cite{snyder}:
$$[x,y] = (\imath a^2/\hbar )L_z, [t,x] = (\imath a^2/\hbar c)M_x, etc.$$
\begin{equation}
[x,p_x] = \imath \hbar [1 + (a/\hbar )^2 p^2_x];\label{e3}
\end{equation}
The relations (\ref{e3}) are compatible with Special Relativity. The first of the relations (\ref{e3}) can be written, for simplicity as:
\begin{equation}
[x,y] \sim 0(l^2) etc.\cdots\label{e4}
\end{equation}
In (\ref{e4}), if we neglect terms of the $\sim l^2, l$ being the minimum interval, typically the Compton (including the Planck) scale, then we return to the usual spacetime and the usual Quantum Theory.\\
The crux is this ``Quantum of Area'', which has emerged as an all important irreduceable unit in the latest theories \cite{baez}. Let us now consider the implications of this minimum area in the usual non-Abelian gauge theory. This will also show how we go beyond Standard Physics.\\
As is well known we consider a generalization of the usual phase function $\lambda$ to include fields with internal degrees of freedom \cite{moriato}. For example $\lambda$ could be replaced by $A_\mu$ to give the Gauge Field 
\begin{equation}
A_\mu = \sum_{\imath} A^\imath_\mu (x)L_\imath ,\label{e5}
\end{equation}
The Gauge Field itself would be obtained by a very well known procedure: 
\begin{equation}
F_{\mu \nu} = \partial_\mu A_\nu - \partial_\nu A_\mu - \imath q [A_\mu , A_\nu ],\label{e6}
\end{equation}
$q$ being the Gauge Field coupling constant.\\
In (\ref{e6}), the second term on the right side is typical of a non Abelian Gauge Field. As is well known, in a typical Lagrangian like 
\begin{equation}
\mathit{L} = \imath \bar \psi \gamma^\mu D_\mu \psi - \frac{1}{4} F^{\mu \nu} F_{\mu \nu} - m \bar \psi \psi\label{e7}
\end{equation}
$D$ denoting the Gauge covariant derivative, there is no mass term for the field Bosons. Such a mass term in (\ref{e7}) must have the form $m^2 A^\mu A_\mu$ which unfortunately is not Gauge invariant.\\
To generate massive gauge bosons, in analogy with superconductivity theory, an extra phase of a self coherent system (Cf.ref.\cite{moriato} for a simple and elegant treatment) has been introduced, as is well known. Thus instead of the Gauge Field $A_\mu$, we consider a new phase adjusted Gauge Field after the symmetry is broken
\begin{equation}
W_\mu = A_\mu - \frac{1}{q} \partial_\mu \phi\label{e8}
\end{equation}
The field $W_\mu$ now generates the mass in a self consistent manner via a Higgs mechanism. Infact the kinetic energy term
\begin{equation}
\frac{1}{2} |D_\mu \phi |^2\quad ,\label{e9}
\end{equation}
where $D_\mu$ in (\ref{e9})denotes the Gauge covariant derivative, now becomes
\begin{equation}
|D_\mu \phi_0 |^2 = q^2|W_\mu |^2 |\phi_0 |^2 \, ,\label{e10}
\end{equation}
Equation (\ref{e8}) gives the mass in terms of the ground state $\phi_0$.\\
It must be remembered that the symmetry breaking of the gauge field is a short length scale phenomenon signifying the fact that the field is mediated by particles with large mass. Further the internal symmetry space of the gauge field is broken by an external constraint: the wave function has an intrinsic relative phase factor which is a different function of space time coordinates compared to the phase change necessitated by the minimum coupling requirement for a free particle with the gauge potential. This cannot be achieved for an ordinary point like particle, but a new type of a physical system, like the self coherent system of Superconductivity Theory now interacts with the gauge field. The second or extra term in (\ref{e6}) is effectively an external field, though (\ref{e6}) manifests itself only in a relatively small spatial interval. The $\phi$ or Higgs field in (\ref{e6}), in analogy with the phase function of Cooper pairs of Superconductivity Theory comes with the Landau-Ginzburg potential $V (\phi)$.\\
Let us now consider in the Gauge Field transformation, an additional phase term, $f(x)$, this being a scalar\cite{bgssymm}. In the usual theory such a term can always be gauged away in the $U(1)$ electromagnetic group. However we now consider the new situation of a noncommutative geometry referred to above, 
\begin{equation}
\left[dx^\mu , dx^\nu \right] = \Theta^{\mu \nu} \beta , \beta \sim 0 (l^2)\label{e11}
\end{equation}
where $l^2$ denotes the minimum spacetime cut off. (Cf. also ref.\cite{nc,annales}) (\ref{e11}) is infact Lorentz covariant. Then the $f$ phase factor gives a contribution to the second order in coordinate differentials,
$$\frac{1}{2} \left[\partial_\mu B_\nu - \partial_\nu B_\mu \right] \left[dx^\mu , dx^\nu \right]$$
\begin{equation}
+ \frac{1}{2} \left[\partial_\mu B_\nu + \partial_\nu B_\mu \right] \left[dx^\mu dx^\nu + dx^\nu dx^\mu \right]\label{e12}
\end{equation}
where $B_\mu \equiv \partial_\mu f$.\\
As can be seen from (\ref{e10}) and (\ref{e9}), the new contribution is in the term which contains the commutator of the coordinate differentials, and not in the symmetric second term. Effectively, remembering that $B_\mu$ arises from the scalar phase factor, and not from the non-Abelian Gauge Field, $A_\mu$ is replaced by 
\begin{equation}
A_\mu \to A_\mu + B_\mu = A_\mu + \partial_\mu f\label{e13}
\end{equation}
Comparing (\ref{e13}) with (\ref{e6}) we can immediately see that the effect of noncommutativity is precisely that of providing a new symmetry breaking term to the Gauge Field, a term which does not come from the Gauge Field itself. Being an $0(l^2)$ effect, it manifests itself only at small scales, as required.\\
Effectively, because of (\ref{e13}) we would have, specializing to a spherically symmetric field for simplicity, instead of the usual Maxwell equations in the gauge field context,
\begin{equation}
\vec E \to \vec E - \vec \nabla f = \vec \nabla Q - \vec \nabla f\label{e14}
\end{equation}
So we have for a point Gauge charge, the modified equation
\begin{equation}
\nabla^2 Q = -4 \pi \rho + \lambda (r)\label{e15}
\end{equation}
The solution of (\ref{e15}) is
\begin{equation}
Q = \int_v \frac{(\rho + \lambda (r)}{r})\label{e16}
\end{equation}
In (\ref{e15}) and (\ref{e16}) $\lambda (r)$ represents the effect of the noncommutativity and is an order of $l^2$ effect, that is it falls off rapidly. It can be seen that the first term in the integral on the right side of (\ref{e16}) gives, in conjunction with (\ref{e14}) the usual Coulumb type of a field. It is the second term in the integral which represents a field due to the noncommutativity of spacetime, which falls off rapidly, as it vanishes at scales where order of $l^2$ can be neglected. As such it represents a field mediated by massive particles.\\
(This is a well known example from the early days of Yang-Mills Theory, which lead to the conclusion that there was a Coulumb type potential of electromagnetism, that is a field without any mass.)\\
It may be remarked that a similar argument using equations like (\ref{e11}) and (\ref{e12}) has been used by the author to argue that one could obtain a reconciliation of electromagnetism and gravitation \cite{annales}.
\section{Cosmology}
We consider a Cosmology in which given the well known $N \sim 10^{80}$ elementary particles, typically pions in the universe, it follows that the pions can be thought of as being created from a Quantum vaccuum, in a phase transition by $n \sim 10^{40}$ Planck particles on the lines of the Prigogine Cosmology (Cf.ref.\cite{prigogine,psp,psu}). Infact let us think of the unit Quanta of Area. In the Quantum vaccuum as being oriented randomly, that is their normals being randomly distributed focussing our attention on the normals, we have a situation similar to the Ising Model. Furthermore considering the amplitudes of these elementary elements of the Quantum vaccuum, we have (Cf.ref.\cite{cu}) a non linear Schrodinger equation,
\begin{equation}
\imath \hbar \frac{\partial \psi}{\partial t} =
\frac{-\hbar^2}{2m'}\frac{\partial^2 \psi}{\partial x^2} + \int
\psi^* (x')\psi (x)\psi (x')U(x')dx',\label{e17}
\end{equation}
(\ref{e17}) is the complete analogue of the Landau-Ginsburg equation
\begin{equation}
-\frac{h^2}{2m} \nabla^2 \psi + \beta |\psi |^2 \psi = -\propto
\psi\label{e16a}
\end{equation}
The correlation loength from (\ref{e16a}) is given by
\begin{equation}
\xi = (\frac{\gamma}{\propto})^{\frac{1}{2}} (\gamma \equiv \hbar^2/2m)\label{e17a}
\end{equation}
It can be seen from (\ref{e17a}) that this is just the Compton length. In other words the Schrodinger equation (\ref{e17}) describes a Landau-Ginsburg like phase transition, as in the Ising Model, and the normals get oriented.\\
As is well known the interesting aspects of this critical point theory (Cf.ref.\cite{good})are universality and scale. Broadly, this
means that diverse physical phenomena follow the same route at the
critical point, on the one hand, and on the other this can happen
at different scales, as exemplified for example, by the course
graining techniques of the Renormalization Group. To highlight
this point we note that in critical point phenomena we have the
reduced order parameter $\bar Q$ and the reduced correlation
length $\bar \xi$. Near the critical point we have relations
like 
$$(\bar Q) = |t|^\beta , (\bar \xi) = |t|^{-\nu}$$
Whence
\begin{equation}
\bar Q^\nu = \bar \xi^\beta\label{e18}
\end{equation}
In (\ref{e18}) typically $\nu \approx 2\beta$. As $\sqrt{Q} \sim
\frac{1}{\sqrt{n}}$ because $\sqrt{n}$ particles are created
fluctuationally, given $N$ particles, and in view of the fractal
two dimensionality of the path
$$\bar Q \sim \frac{1}{\sqrt{n}}, \bar \xi = (l_P/l)^2$$
This gives
\begin{equation}
l = \sqrt{n}l_P\label{e19}
\end{equation}
In the above phase transition in which the Planck oscillators and elementary particles are created, given $N$ particles at any stage, $\sqrt{N}$ particles are fluctuationally created in the minimum spacetime cut off intervals $l, \tau$, rather on the lines of Hayakawa \cite{hayakawa}. Indeed we will see that this is related to the Modified Uncertainty Principle which itself arises due to the minimum spacetime cut off.\\
To proceed, it was shown in the above references that
\begin{equation}
M = \sqrt{n} m_P, \, m = \sqrt{n'} m_P\label{eA}
\end{equation}
where $M$ and $m$ denote the mass of the universe and the mass of the pion and $m_P$ the Planck mass.
In the following we will use $N$ as the sole cosmological parameter.\\
Equating the gravitational potential energy of the pion in a three dimensional isotropic sphere of pions of radius $R$, the radius of the universe, with the rest energy of the pion, we can deduce the well known relation \cite{nottale,hayakawa}
\begin{equation}
R \approx \frac{GM}{c^2}\label{e16b}
\end{equation}
where $M$ can be obtained from (\ref{eA}).\\
We now use the fluctuation in the particle number of the order $\sqrt{N}$ \cite{hayakawa,ijmpa,ijtp} while a typical time interval for the fluctuations is $\sim h/mc^2$, the Planck (or Compton) time. We will come back to this point later but as mentioned this is also related to the Modified Uncertainty Principle which gives an extra (or duality) term to the usual Heisenberg relation. So we have
$$\frac{dn}{dt} = \frac{\sqrt{n}}{\tau_P}$$
with a similar equation for $N$. Whence on integration we get,
\begin{equation}
T = \left(\hbar/mpc^2\right) \sqrt{n} = \frac{\hbar}{mc^2} \sqrt{N}\label{e17b}
\end{equation}
We can easily verify that equation (\ref{e17b}) is indeed satisfied where $T$ is the age of the universe. Next by differentiating (\ref{e16b}) with respect to $t$ and further steps we get
\begin{equation}
\frac{dR}{dt} \approx HR\label{e18b}
\end{equation}
where $H$ in (\ref{e18b}) can be identified with the Hubble Constant, and using (\ref{eA}) and (\ref{e16b}) is given by,
\begin{equation}
H = Gm_P/(c^2\tau_P R) = \frac{Gm^3c}{\hbar^2}\label{e19b}
\end{equation}
Equation (\ref{e16b}) and (\ref{e17b}) show that in this formulation, the correct mass, radius and age of the universe can be deduced given $N$ as the sole cosmological or large scale parameter Equation (\ref{e19b}) can be written as
\begin{equation}
m \approx \left(\frac{H\hbar^2}{Gc}\right)^{\frac{1}{2}}\label{e20}
\end{equation}
Equation (\ref{e20}) has been empirically known as an ``accidental'' or ``mysterious'' relation. As observed by Weinberg \cite{weinberg}, this is unexplained: it relates a single cosmological parameter $H$ to constants from microphysics. We will touch upon this micro-macro nexus again. In our formulation, equation (\ref{e20}) is no longer a mysterious coincidence but rather a consequence.\\
As (\ref{e19}) and (\ref{e18b}) are not exact equations but rather, order of magnitude relations, it follows that a small cosmological constant $\wedge$ is allowed such that
$$\wedge \leq 0 (H^2)$$
This is consistent with observatioins and shows that $\wedge$ is very very small - this has been a puzzle, the so called cosmological constant problem in the earlier theory \cite{wein}. But it is explained here.\\
To proceed we observe that because of the fluctuation of $\sim \sqrt{N}$ (due to the ZPF), there is an excess electrical potential energy of the electron, which infact we have identified as its inertial energy. That is \cite{ijtp,hayakawa},
$$\sqrt{N} e^2 /R \approx mc^2$$
On using (\ref{e16b}) in the above, we recover the well known Gravitation-electromagnetism ratio viz.,
\begin{equation}
e^2/Gm^2 \sim \sqrt{N} \approx 10^{40}\label{e21}
\end{equation}
or without using (\ref{e16b}), we get, instead, the well known so called Eddington formula,
\begin{equation}
R = \sqrt{N} l \, \mbox{or} \, R = \sqrt{n} l_P\label{e22}
\end{equation}
Infact (\ref{e22}) is the spatial counterpart of (\ref{e17}). If we combine (\ref{e22}) and (\ref{e16b}), we get,
\begin{equation}
\frac{Gm}{lc^2} = \frac{1}{\sqrt{N}} \propto T^{-1}\label{e23}
\end{equation}
where in (\ref{e23}), we have used (\ref{e17b}). Following Dirac (cf. also \cite{melnikov} we get $G$ as the variable, rather than the quantities $m,l,c$ and $\hbar$ (which we will call microphysical constants) because of their central role in atomic (and sub atomic) physics.\\
Next if we use $G$ from (\ref{e23}) in (\ref{e19b}), we can see that
\begin{equation}
H = \frac{c}{l} \frac{1}{\sqrt{N}} = \frac{c}{l_P} \cdot \frac{1}{\sqrt{n}}\label{e24}
\end{equation}
Thus apart from the fact that $H$ has the same inverse time dependance on $T$ as $G$, (\ref{e24}) shows that given the microphysical constants, and $N$, we can deduce the Hubble Constant also as from (\ref{e24}) or (\ref{e19b}).\\
Using (\ref{e16b}), we can now deduce that
\begin{equation}
\rho \approx \frac{m}{l^3} \frac{1}{\sqrt{N}}\label{e25}
\end{equation}
Next (\ref{e22}) and (\ref{e17b}) give,
\begin{equation}
R = cT\label{e26}
\end{equation}
(\ref{e25}) and (\ref{e26}) are consistent with observation.\\
The above model predicts a dark energy driven ever expanding and accelerating universe whose density keeps decreasing. This seemed to go against the accepted idea that the density of the universe equalled the critical density required for closure. But the work of Perlmutter and others as also observations from the Wilkinson Microwave Anisotropy Probe and the Sloan Digital Sky Survey has confirmed the view \cite{science}.
\section{Discussion}
We will now argue that gravitation can be considered to be a residual effect which are on the lines of the Sakharov Model. As noted, it was shown that the pion and the universe itself could be thought of as being made up of Planck oscillators (Cf.ref.\cite{psp,psu}) infact denoting a typical frequency by $\omega$ where
$$\omega = mc^2/\hbar$$
we have
\begin{equation}
\omega = \omega_P l_P/l\label{e27}
\end{equation}
where the subscript $P$ denotes the Planck scale and $l$ would be for the pion, its Compton scale and for the universe itself it would denote $R$. For the pion (\ref{e27}) gives the pion mass $m$, which shows that the pion is the lowest energy state of $\sim 10^{40}$ Planck oscillators. For the universe with $n' \sim 10^{120}$ Planck oscillators, $l$ on the right side of (\ref{e27}) would be the radius $R$ and then the left side would yield the lowest energy state in this case. The highest energy state of $n'$ oscillators would then be, $n' \omega$ which on using (\ref{e27}) yields the correct mass of the universe. Moreover the inverse dependance on distance, of the energy, in (\ref{e27}) indicates that the energy would also be characterized as the potential energy of an inverse square interaction, for example the gravitational interaction. In that case, $R = \frac{GN}{c^2}$ which on using (\ref{e27}) gives for the equation
\begin{equation}
G = lc^2/m \sqrt{N} = \frac{l\hbar}{m^2T} \equiv \frac{\gamma}{T}\label{e28}
\end{equation}
What this means is that without taking recourse to gravitation in the first instance and using the fact that the energy is that of the underpinning of normal mode of Planck oscillators, it then follows that gravitation shows up as a manifestation of this energy, distributed over $N$ particles of the universe. Thus gravitation is now reduced to the status of a statistical measure of residual energies as in Sakharov's formulation using the Planck scale \cite{mwt} as is confirmed by (\ref{e21}).\\
2. In the Bekenstelin Black Hole Radiation Formula if we introduce for $G$, it is time dependent version (\ref{e28}) then we get for the radiation time, instead of the usual expression
$$T /propto M^3$$
$M$ being the mass of the Schwarschild Black Hole, this time
$$\Theta^{-3} T^3 \equiv T^3 \left(\frac{\hbar c^4}{(30.8)^3 \pi \gamma^2}\right) = M^3$$
whence we have
\begin{equation}
T = \Theta M\label{e29}
\end{equation}
One can easily verify that (\ref{e29}) gives the Planck time for a Planck mass and $T \sim 10^{17}$ seconds, the age of the universe, for $M$ the mass of the universe itself. Alternatively (\ref{e29}) can be written as
$$T/t_P = M/m_P$$
which is also easily verified to be true.\\
3. Another way of interpreting the result in the previous section is that the area of a pion is $n$ times the elemental Planck area as in the Quantum Gravity description, and similarly the area of the universe is $N$ times the elementary particle area. That the area shows up as being fundamental can be interpreted as due to the two dimensionality of the Brownian Quantum path in (\ref{e19}), now viewed as a random walk equation as discussed in \cite{psp,psu}.\\
4. Finally, it may be remarked that the above characterization of $G$, not only reproduces standard results like the bending of light or the precession of the perihelion of Mercury, but also provides an explanation for the otherwise inexplicable anomalous accelerations of the Pioneer spacecrafts \cite{nc115,csf}. 


\begin{thebibliography}{99}
\bibitem {mwt} C.W. Misner, K.S. Thorne and J.A. Wheeler, ``Gravitation'', W.H. Freeman, San Francisco, 1973, pp.819ff.
\bibitem {gerard} G 't Hooft in Proceedings of Fourth International Symposium ``Froniers of Fundamental Physics 4'', Kluwer Academic, New York, 2001, 1-12.
\bibitem {hooft} G 't Hooft, arXiv:gr-qc/9305008.
\bibitem {cu} B.G. Sidharth, "Chaotic Universe: From the Planck to the Hubble Scale", Nova Science Publishers, Inc., New York, 2001.
\bibitem {dirac} P.A.M. Dirac, ``The Principles of Quantum Mechanics'', Clarendon Press, Oxford, 1958, p.263.
\bibitem {snyder} H.S. Snyder, Physical Review, Vol.72, No.1, July 1 1947, p.68-71.
\bibitem {baez} J. Baez, Nature, 421, February 2003, pg.702.
\bibitem {moriato} K. Moriyasu, ``An Elementary Primer for Gauge Theory'', World Scientific, Singapore, 1983.
\bibitem {bgssymm} B.G. Sidharth, Symmetry Breaking in Gauge Field Theory Due to Noncommutative Spacetime'', to appear in Int.J.Mod.Phys.A.
\bibitem {nc} B.G. Sidharth, Il Nuovo Cimento, 117B (6), 2002, 703ff.
\bibitem {annales} B.G. Sidharth, Gravitation from a Gauge like Formulation, to appear in Annales de la Fondation Louis de Broglie.
\bibitem {prigogine} G.Nicolis and G. Prigogine in ``Exploring Complexity'', W.H. Freeman, New York, 1989, p.10.
\bibitem {psp}  B.G. Sidharth, Found.Phys.Letters., 15 (6), 2002, 577-583.
\bibitem {psu} B.G. Sidhartyh, Found.Phys.Letters., 17 (5), 2004, 503-506.
\bibitem {good} D.L. Goodstein, "States of Matter", Dover Publications, Inc., New York, 1975, p.462ff.
\bibitem {hayakawa} S. Hayakawa, Suppl of PTP Commemmorative Issue, 1965, 532-541.
\bibitem {nottale} L. Nottale, ``Fractal Space-Time and Microphysics:Towards a Theory of Scale Relativity'', World Scientific, Singapore, 1993, p.312.
\bibitem {ijmpa} B.G. Sidharth, Int.J.Mod.Phys.A., {\bf 13} (15), 1993,pp.2599ff.
\bibitem {ijtp} B.G. Sidharth, Int.J.Th.Phys., Vol.37, No.4, 1998, 1307-1312.
\bibitem {weinberg} S. Weinberg, ``Gravitation and Cosmology'', John Wiley \& Sons, New York, 1972, p.62.
\bibitem {wein} S. Weinberg, Phys.Rev.Lett., 43, 1979, p.1566.
\bibitem {melnikov} V.N. Melnikov, Int.J.Th.Phys., \underline{33} (7), 1994, 1569-1579.
\bibitem {science} Science, 302 (5653); 2003, 2038.
\bibitem {nc115} B.G. Sidharth, Nuovo Cimento 115B, 12, 2000, pp.151ff.
\bibitem {csf} B.G. Sidharth, 22 (3), 2004, pp.537ff.
\end{thebibliography}
\end{document}